\documentclass[twocolumn,aps,prb,superscriptaddress,showpacs,floatfix,amsmath]{revtex4-2}
\usepackage{epsfig}
\usepackage{amsmath}
\usepackage{bm}
\usepackage{here}
\usepackage{float}
\usepackage{graphicx}
\usepackage[linkcolor=black,urlcolor=blue,citecolor=blue,colorlinks=true]{hyperref}
\usepackage{xcolor}

\usepackage[normalem]{ulem}

\begin{document}
	
\title{Layer-resolved electronic behavior in a Kondo lattice system, CeAgAs$_2$}
	
\author{Sawani Datta,$^1$ Ram Prakash Pandeya,$^1$ Arka Bikash Dey,$^2$ A. Gloskovskii,$^2$ C. Schlueter,$^2$ T. R. F. Peixoto,$^2$ Ankita Singh,$^1$ A. Thamizhavel,$^1$ and Kalobaran Maiti}
\altaffiliation{Corresponding author: kbmaiti@tifr.res.in}
\affiliation{Department of Condensed Matter Physics and Materials Science, Tata Institute of Fundamental Research, Homi Bhabha Road, Colaba, Mumbai-400005, India.\\
$^2$Deutsches Elektronen-Synchrotron DESY, 22607 Hamburg, Germany.}
	
	
\begin{abstract}

We investigate the electronic structure of an antiferromagnetic Kondo lattice system CeAgAs$_2$ employing hard $x$-ray photoemission spectroscopy. CeAgAs$_2$, an orthorhombic variant of HfCuSi$_2$ structure, exhibits antiferromagnetic ground state, Kondo like resistivity upturn and compensation of magnetic moments at low temperatures. The photoemission spectra obtained at different photon energies suggest termination of the cleaved surface at cis-trans-As layers. The depth-resolved data show significant surface-bulk differences in the As and Ce core level spectra. The As 2$p$ bulk spectrum shows distinct two peaks corresponding to two different As layers. The peak at higher binding energy correspond to cis-trans-As layers and is weakly hybridized with the adjacent Ce layers. The As layers between Ce and Ag-layers possess close to trivalent configuration due to strong hybridization with the neighboring atoms and the corresponding feature appear at lower binding energy. Ce 3$d$ core level spectra show multiple features reflecting strong Ce-As hybridization and strong correlation. Intense $f_0$ peak is observed in the surface spectrum while it is insignificant in the bulk. In addition, we observe a features at binding energy lower than the well-screened feature indicating the presence of additional interactions. This feature becomes more intense in the bulk spectra suggesting it to be a bulk property. Increase in temperature leads to a spectral weight transfer to higher binding energies in the core level spectra and a depletion of spectral intensity at the Fermi level as expected in a Kondo material. These results reveal interesting surface-bulk differences, complex interplay of intra- and inter-layer covalency, and electron correlation in the electronic structure of this novel Kondo lattice system.
		
\end{abstract}
	
\maketitle
	
\section{Introduction}

\begin{figure}
\centering
\includegraphics[width=0.5\textwidth]{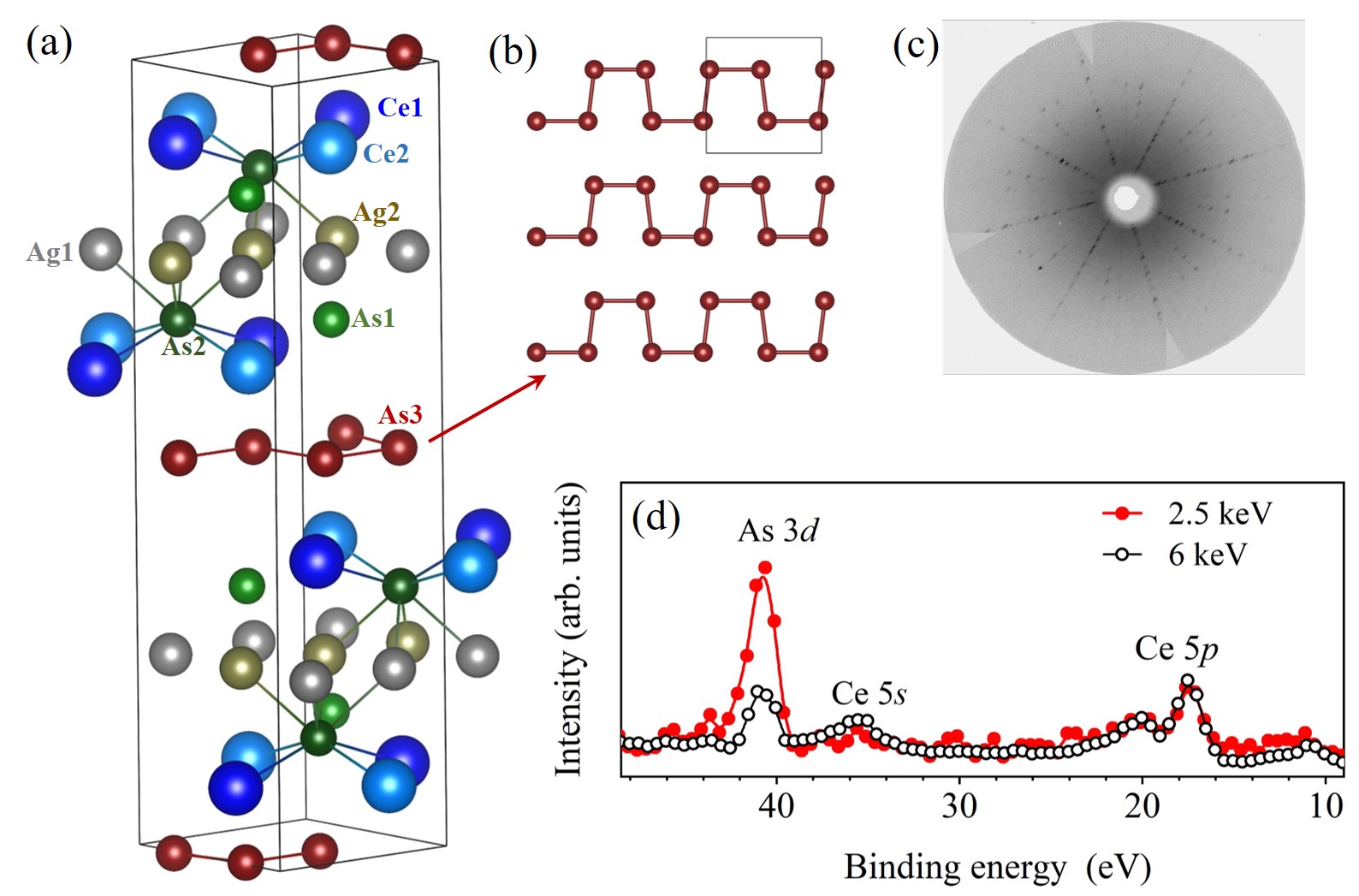}
\caption{(a) Crystal structure of CeAgAs$_2$. (b) Cis-trans-chain arrangement of As3 atoms in the $xy$-plane. (c) Laue pattern, exhibiting the high-quality of the single crystalline sample. (d) As 3$d$, Ce 5$s$ and Ce 5$p$ core-level data collected using 2.5 keV (open circles) and 6 keV (closed circles) photon energies.}
\label{structure}
\end{figure}

The 112-type intermetallics, (A/RE)TX$_2$ (A/RE = alkali/rare earth elements, T = transition metal, X = pnictogens) have been studied for many decades due to their large varieties of interesting physical properties such as different types of magnetic ordering (ferromagnetic\cite{Liu, thamizh}, antiferromagnetic \cite{thamizh, seibel}), signature of superconductivity\cite{Han}, anomalous low-temperature behaviour\cite{Sengupta1, Sengupta2}, etc. For the past few years, this group of samples has attracted additional attention involving exotic non-trivial topological states\cite{ruszala}. In this group, the CeTX$_2$ compounds are of particular interest due to the presence of electron correlation in the electronic structure induced by the Ce 4$f$ states appearing near Fermi level, $\epsilon_F$\cite{sawani,chainani,CeCuSb2}. These materials crystallize in the tetragonal HfCuSi$_2$ type structure belonging to the space group, $P4/nmm$\cite{layer_brylak, USb2_Kac}. This type of structure consists of a pnictogen layer with a pseudo-2 dimensional (2D) square net arrangement followed by a tightly packed Ce-X-T-X-Ce layers. While the pnictogen atoms forming the square-net structure contribute to the topological non-trivial states leading to enormous attraction in the field\cite{sebastian}, Peierls-like distortions are often observed in these materials which breaks the square-net structure \cite{Tremel}. One example of this type of distortion in the pnictogen square-net has been observed in CeAgAs$_2$; the crystal structure is shown in Fig. \ref{structure}(a). In this structure, Ce-As-Ag-As-Ce layers are bonded together. As3 atoms form a cis-trans type chains instead of the square-net structure as shown in Fig. \ref{structure}(b). This deformation breaks the tetragonal symmetry of the system and follows the orthorhombic symmetry belonging to the $Pmca$ space group\cite{demchyna}.

CeAgAs$_2$ exhibits two antiferromagnetic transitions with N\'{e}el temperatures, 6 K and 4.9 K \cite{rajib, Maria}. Transport measurements show negative logarithmic increase of the electrical resistivity below 25 K suggesting a Kondo-like behavior. The effective magnetic moment, $\mu_{eff}$ estimated in the paramagnetic regime is 2.59 $\mu_B$/Ce and 2.66 $\mu_B$/Ce for the applied external magnetic field, $H||$~[100] and $H||$~[010] directions, respectively\cite{rajib}; $\mu_{eff}$ is close to the expected value for trivalent Ce. The magnetic moment reduces significantly at low temperatures;  the saturation magnetisation is 1.2 $\mu_B$/Ce for $H||$~[100] at 2 K\cite{rajib}. Such large compensation of the local moment is consistent with the Kondo-type transport behavior \cite{swapnil}. Here, we investigated the electronic structure of CeAgAs$_2$ using hard $x$-ray photoemission spectroscopy (HAXPES). The experiments performed using varied photon energies and temperatures reveal interesting spectral renormalization.

\section{Method}
	
High-quality single crystals of CeAgAs$_2$ were prepared by flux method using Ag-As as flux \cite{rajib}. Chemical composition of the sample was verified via the energy dispersive analysis of $x$-rays. Single crystallinity of the sample was ensured by Laue diffraction measurements; a typical Laue diffraction pattern is shown in Fig. \ref{structure}(c) exhibiting sharp bright spots manifesting excellent crystallinity. Electronic structure of CeAgAs$_2$ was calculated following the density functional theory (DFT) using the full potential linearized augmented plane wave method (FLAPW) as captured in the Wien2k software\cite{Blaha}. These density functional theory (DFT) calculations\cite{Hohen} were carried out using the Perdew-Burke-Ernzerhof generalized gradient approximation (GGA)\cite{Perdew} including spin-orbit coupling and effective electron correlation strength, $U_{eff}$ = 8 eV for the Ce 4$f$ electrons.

The HAXPES measurements were carried out at the P22 beamline of PETRA III, DESY, Hamburg, Germany, using a high-resolution Phoibos electron analyzer with the energy resolution of 150 meV for 6 keV photon energies. The resolution at other photon energies was 200 meV. All the measurements were carried out on sample surfaces freshly cleaved in ultrahigh vacuum (pressure $\sim$ 10$^{-10}$ Torr). During the measurements, the chamber was maintained in the ultra high vacuum condition ($<$ 10$^{-10}$ Torr). The sample temperature was varied using an open cycle He cryostat. To probe different depths of the sample, we have used three widely separated photon energies 2.5, 6, and 8 keV \cite{escapeDepth}.
	
\section{Results and discussions}
	
The unit cell of CeAgAs$_2$, shown in Fig. \ref{structure}(a), is formed by periodic stacking of  [As3]-[(Ce1/Ce2)-(As1/As2)-(Ag1/Ag2)-(As1/As2)-(Ce1/Ce2)] layers. In HfCuSi$_2$ structure, As3 layers form a planar square-net structure. However, in CeAgAs$_2$, As3 layers form cis-trans-chains in the $xy$-plane \cite{demchyna} as shown in Fig. \ref{structure}(b). As1/As2 atoms form a slightly distorted tetragonal anti-prismatic structure with Ce and Ag atoms. The lattice parameters of CeAgAs$_2$ are: $a$ = 5.7586 \AA, $b$ = 5.7852 \AA, and $c$ = 21.066 \AA \cite{escapeDepth}. The separation of As3 layers from other layers is large with weak inter-layer coupling. Therefore, it is likely that the cleaved surface will be either As3 layer or Ce layer. In order to identify terminated layer experimentally, we studied the shallow core level (Ce 5$p$ and As 3$d$) data collected using 2.5 keV and 6 keV photon energies as shown in Fig. \ref{structure}(d). The photoelectron escape depth is about 26 \AA\ at 2.5 keV and 40 \AA\ at 6 keV photon energy. Such change in probing-depth leads to a significant change in intensity of the experimental features; spectra are normalized by the intensity of the Ce 5$p$ features. It is evident that the relative intensity of As 3$d$ peak becomes much stronger in the surface sensitive 2.5 keV data (the ratio of the photoemission cross-section of As 3$d$ and Ce 5$p$ core levels are similar in these two photon energies). This suggests that surface electronic structure of the cleaved sample studied here is dominantly contributed by the As layers and Ce-layers appear underneath the As-layer. Later in the text we will establish based on the experimental results that As3-layer is the surface layer.

\begin{figure}
\centering
\includegraphics[width=0.5\textwidth]{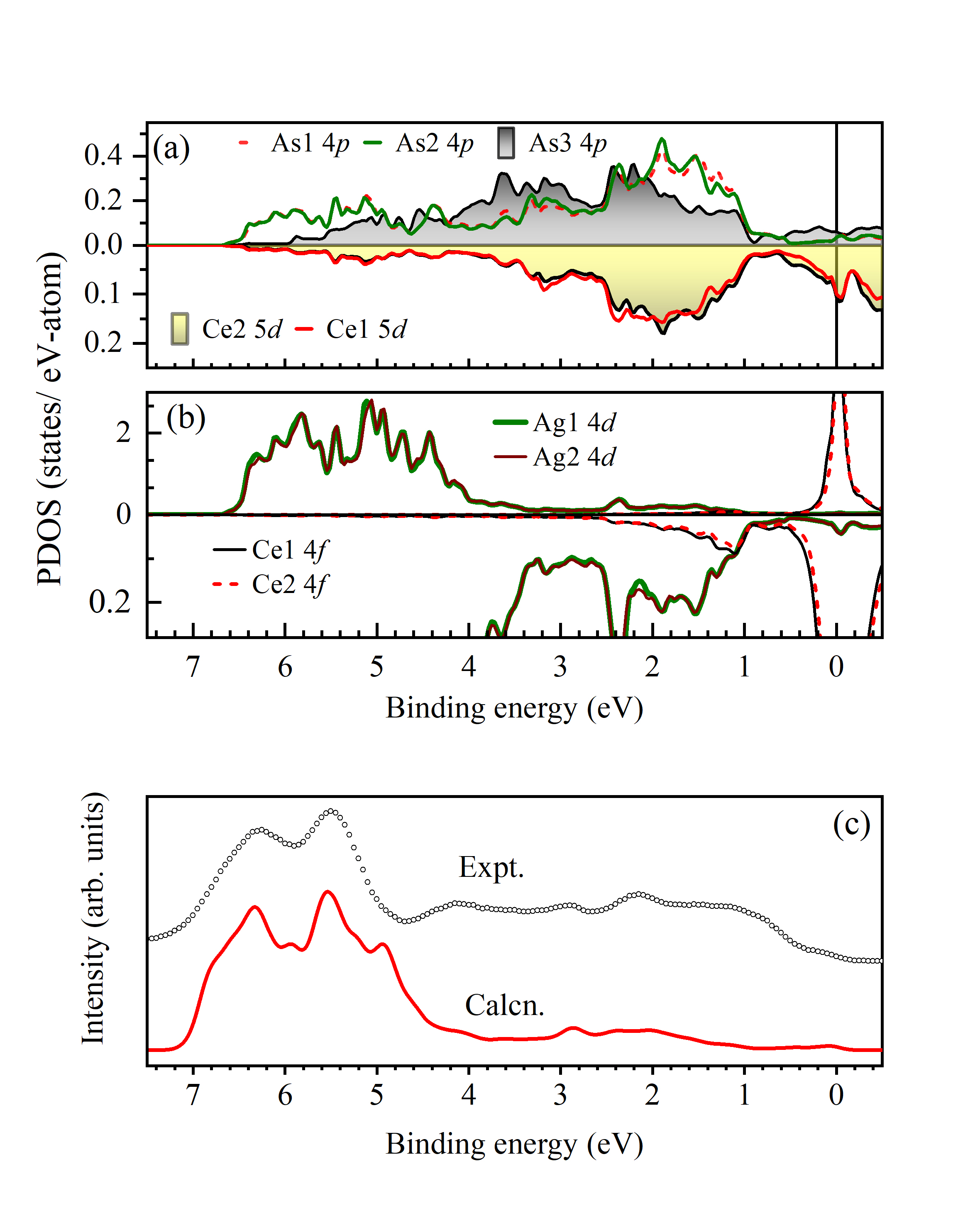}
\vspace{-4ex}
\caption{Calculated partial density of states (PDOS) of (a) As 4$p$ (upper panel) and Ce 5$d$ (lower panel) contributions. (b) Ag 4$d$, Ce 4$f$ contributions (upper panel). Ag 4$d$ and Ce 4$f$ PDOS in an expanded PDOS scale is shown in the lower panel. (c) Experimental valence band spectra (symbols) collected using 6 keV photon energy at 45 K and calculated total density of states.}
\label{DFT}
\end{figure}

The calculated partial density of states (PDOS) of CeAgAs$_2$ are shown in Fig. \ref{DFT}. As1/As2 4$p$ PDOS in Fig. \ref{DFT}(a) appear very similar and exhibit signatures of bonding and antibonding bands around the binding energies, 5.5 eV and 2 eV, respectively. Ce 5$d$ PDOS shown in the lower panel of Fig. \ref{DFT}(a) and Ag 4$d$ PDOS in the upper panel of Fig. \ref{DFT}(b) exhibit energy distribution similar to As1/As2 4$p$ PDOS with dominant Ag contributions in the higher binding energy regime while Ce 5$d$ PDOS is more intense near $\epsilon_F$. These results indicate that As1/As2 are strongly bonded to Ce and Ag atoms.

As3 4$p$ PDOS appear in the binding energy range of 1-6 eV. As3 4$p$ intensity at $\epsilon_F$ is small exhibiting a dip in PDOS and is similar to the As1/As2 contributions exhibiting a small peak like structure. This scenario is different from the square net structured materials in the CeTX$_2$ family where the square net As 4$p$ contributions at $\epsilon_F$ is larger than other As contributions\cite{ruszala}. The distortion in the square-net leads to a transfer of As3 4$p$ contributions away from $\epsilon_F$ lowering the total energy of the system\cite{Maria}. The energy distribution of As3 4$p$ PDOS is different from the PDOS distributions of As1/As2, Ce and Ag valence states reflecting weak hybridization between them consistent with the behavior expected from the structural parameters. Ce 4$f$ PDOS are shown in the upper panel of Fig. \ref{DFT}(b); the same is shown in an expanded scale in the lower panel along with Ag 4$d$ PDOS. The hybridization of Ce 4$f$ states with the conduction electronic states consisting of Ce 5$d$, As 4$p$ and Ag 4$d$ states is evident leading to significant intensities between 1-2 eV binding energies at $\epsilon_F$.

In Fig. \ref{DFT}(c), we show the experimental valence band spectrum (open circles) obtained at 45 K using 6 keV photon energy exhibiting multiple features. Due to the large escape depth of about 40 \AA\ of the valence electrons at 6 keV photon energy, the valence band spectra predominantly represent the bulk contributions ($>$84\%). Therefore, we investigate below the experimental data considering the results from the bulk calculations.  The peaks between 5-7 eV binding energies correspond to the photoemission from the energy bands having dominant Ag 4$d$ contributions. The intensities in the 0-5 eV binding energies appear due to As 4$p$, Ce 5$d$ and Ce 4$f$ contributions which corroborate well with the calculated results. The comparison with the DFT results suggests that the intensities in the energy region 2.5 - 5 eV binding energies corresponds primarily to the non-bonding As 4$p$ states. The intensities around 2 eV are the contributions from the antibonding bands corresponding to Ag 3$d$ - As 4$p$ hybridization and bonding states corresponding to As 4$p$ - Ce 5$d$4$f$ hybridizations; these intensities have dominant As 4$p$ character as evident from the calculated PDOS. Considering that atomic photoemission cross-sections\cite{Yeh} provide a reasonable approximation of the photoemission from various PDOS, we calculated the spectral functions of all the constituent PDOS, convoluted with the Fermi-Dirac distribution function and broadened by the energy resolution function. Ag 4$d$ PDOS is shifted by 0.5 eV towards higher binding energy to match the Ag 4$d$ peak positions in the experimental data; such rigid energy shift often works well to account for the electron correlation induced effect of fully filled and/or close to fully filled bands. The calculated total spectral function is shown by solid line in Fig. \ref{DFT}(c). Clearly, the features in the experimental spectra could be captured excellently well in the calculated spectral function though the relative intensities in some cases are different due to the consideration of the atomic cross-section.

\begin{figure}
\centering
\includegraphics[width=0.5\textwidth]{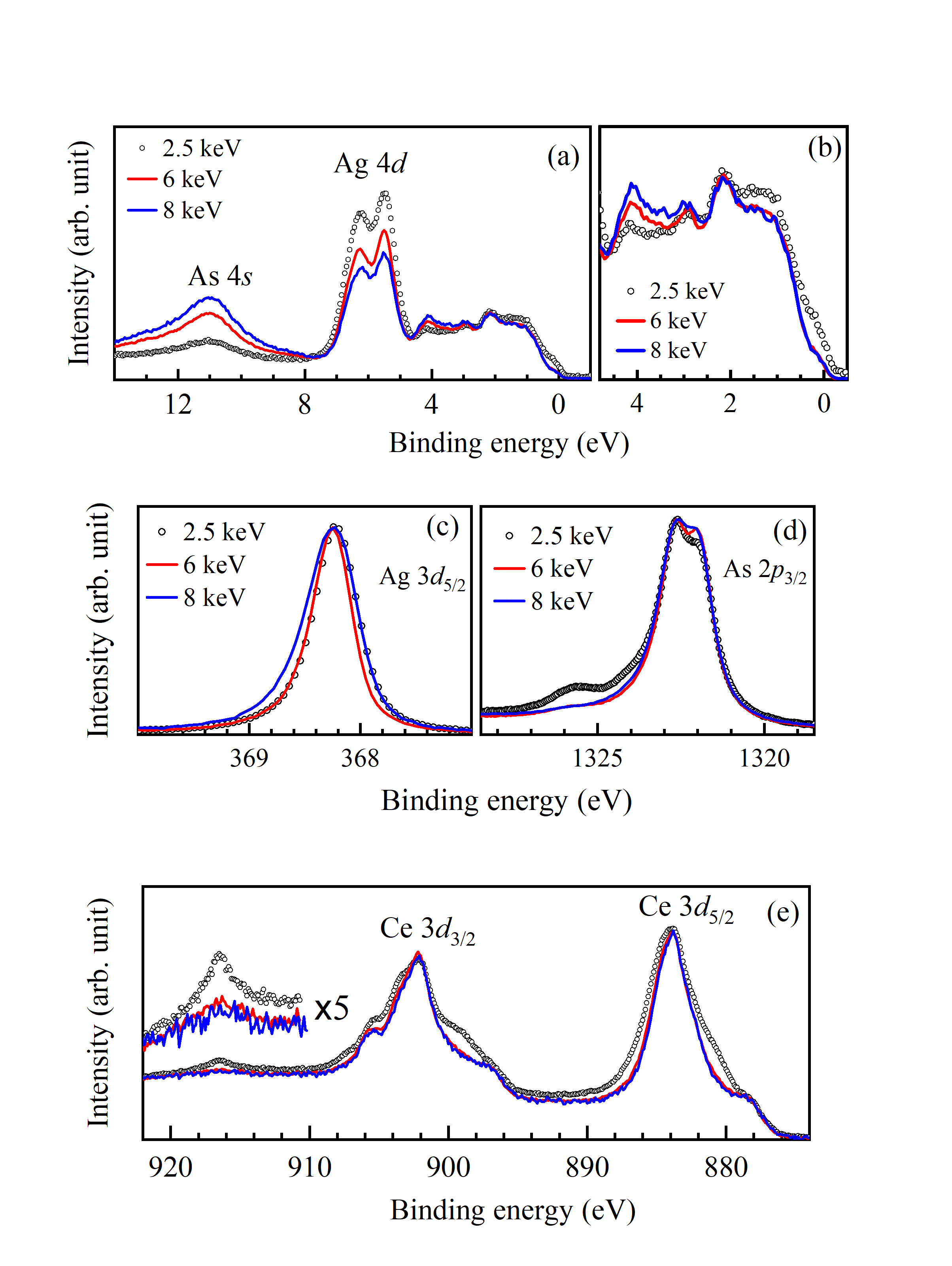}
\vspace{-4ex}
\caption{(a) Valence band spectra collected using 2.5 keV, 6 keV and 8 keV photon energies at 45 K. (b) Near $\epsilon_F$ region of the valence band in an expanded scale for clarity. (c) Ag 3$d_{5/2}$, (d) As 2$p_{3/2}$, and (e) Ce 3$d$ spectra collected using 2.5 keV (open circles), 6 keV (red line) and 8 keV (blue line) photon energies.}
\label{Depth}
\end{figure}

Valence band spectra collected at 45 K using different photon energies are shown in Fig. \ref{Depth}(a). We observe significant change in intensity of the spectral features with the change in photon energy. Understanding of such spectral evolution is difficult due to the contributions from various hybridized states whose photoemission cross-section \cite{Yeh} changes differently with the change in photon energy. The probing depth will change due to the change in photoelectron escape depth, $\lambda$. Light polarization also plays an important role in deriving the intensity of the spectral features \cite{Polarization}. From the photoemission cross-section\cite{Yeh} of various states we note that the cross section of Ag 4$d$ reduces by about three orders of magnitude (230 times) if the photon energy changes from 1486.6 eV to 8047.8 eV. The relative change is much smaller (27 times) for As 4$s$ states. Thus, significant decrease in Ag 4$d$ peaks and increase of As 4$s$ peak in Fig. \ref{Depth}(a) is attributed primarily due to the change in photoemission cross-section; spectra are normalized by the intensities at 2 eV binding energy.

There are significant spectral redistributions in the energy region closer to $\epsilon_F$ due to the change in photon energy which is shown in an expanded scale in Fig. \ref{Depth}(b) for clarity. The energy region 2-4 eV is essentially contributed by As 4$p$ non-bonding states as found from the DFT calculations. The intensity around 4 eV binding energy increases with the increase in photon energy while an opposite trend is seen in the vicinity of $\epsilon_F$. Since Ce 4$f$ cross-section decreases more rapidly with the increase in photon energy than the decrease in cross-section of Ce 5$d$ and As 4$p$ photoemission, the intensities near $\epsilon_F$ can be attributed primarily to the Ce 4$f$ matrix element effect of the photoemission process. The intensities at higher binding energies appear due to Ce 5$d$-As 4$p$ hybridized states. The 4 eV features has dominant As3 4$p$ contributions (see Fig. \ref{DFT}); the intensity increase may be a reflection of matrix element effect.\cite{Yeh} In addition, the surface-bulk difference in the electronic structure will also contribute in the spectral intensity redistribution.

In order to probe the surface-bulk differences in the electronic structure, we study the core level spectra as a function of photon energy. Ag 3$d_{5/2}$ spectra shown in Fig. \ref{Depth}(c) exhibit a change in linewidth with photon energy; the 6 keV data show the narrowest linewidth presumably due to the best energy resolution in this case. The data at 8 keV exhibit larger asymmetry compared to 2.5 keV data although the resolution broadening in these two cases are very similar.  The peak position is almost the same in all the three cases. These results suggest that Ag layers are located in the bulk of the sample or the surface-bulk difference is not significant for Ag contributions. Later we will see that the experiment shows results in favor of the former argument. While the change in linewidth/asymmetry may appear due to the change in resolution and lifetime broadenings, significantly larger asymmetry in the 8 keV case may be due to larger dispersion of the valence band due to relatively stronger hybridization in the bulk.

The As 2$p_{3/2}$ spectra shown in Fig. \ref{Depth}(d) exhibit significant change with the change in photon energy; a new intense feature is observed at around 1326 eV in the 2.5 keV spectrum apart from a change in relative intensities of the features near 1322 eV. Clearly, surface and bulk As 2$p$ must be very different. Ce 3$d$ spectra shown in Fig. \ref{Depth}(e) also exhibit significant modification with the change in photon energy. In all these cases, the data collected with 6 keV and 8 keV photon energies are very similar and represent essentially the bulk electronic structure while the 2.5 keV data are very different due to relatively larger surface contributions. There is an intense peak at 916 eV binding energy which becomes sharp and intense at 2.5 keV as shown in an expanded scale in Fig. \ref{Depth}(e). Large enhancement in the 2.5 keV data suggests that this feature may be associated to the surface electronic structure.

\begin{figure}
\centering
\includegraphics[width=0.5\textwidth]{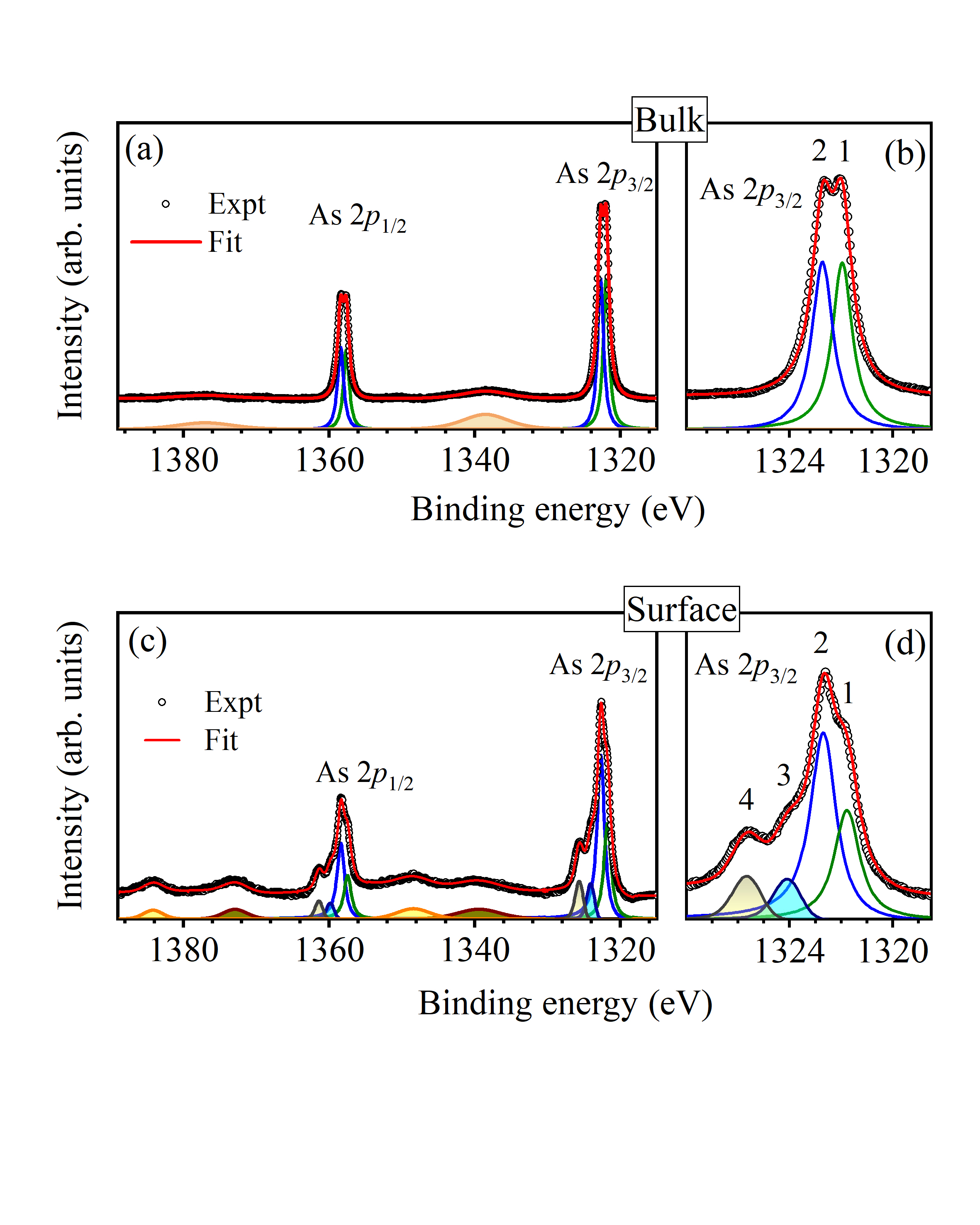}
\vspace{-12ex}
\caption{As 2$p$ (a) bulk and (c) surface spectra. The lines represent the simulated spectra using a Voigt lineshapes for each distinct feature. (b) Bulk and (d) surface As 2$p_{3/2}$ spectra in an expanded energy scale to show the features with better clarity.}
\label{As2pFit}
\end{figure}

The photoemission intensity at a binding energy, $\epsilon$ can be expressed as, $I(\epsilon) = \int_{0}^{d}I^s(\epsilon)e^{-x/\lambda}dx + \int_{d}^{\infty}I^b(\epsilon)e^{-x/\lambda}dx$. Here $I^s(\epsilon)$ and $I^b(\epsilon)$ are the surface and bulk spectral functions; $d$ and $\lambda$ are the effective thickness of the surface layer and the photoelectrons escape depth, respectively. By integrating the above equation we get, $I(\epsilon) = I^s(\epsilon)(1 - e^{-d/\lambda}) + I^b(\epsilon)e^{-d/\lambda}$. Since the number of core electrons are same in the surface and bulk electronic structures, and the core level photoemission spectra represents the electron density in the corresponding energy levels, we normalized the spectral functions by the area under the core level spectra for the extraction of the surface-bulk electronic structure following the well-established method\cite{surface-bulk}. Here, $I^s(\epsilon)$, $I^b(\epsilon)$, $d/\lambda$(2.5 keV) and  $d/\lambda$(6 keV) are the four unknown parameters. The surface and bulk spectral functions, and $d/\lambda$ are calculated using the above equations for 2.5 keV and 6 keV data along with the constraints that $\lambda$ follows the universal curve and the spectral intensity should be positive at all energies (negative intensity is unphysical)\cite{escapeDepth}. The extracted bulk and surface spectra are shown by the open circle in Fig. \ref{As2pFit}(a) and (c) respectively; the expanded part of the 2$p_{3/2}$ region is shown in Fig. \ref{As2pFit}(b) and (d). We have also fitted the extracted spectra by least-square error method using a set of Voigt functions. The simulated curve is shown by the red line superimposed over the experimental data in the figure. The constituent peaks are shown in the lower panel in each case. For the simulation, we have kept the intensity ratio of the spin-obit split peaks fixed to the ratio of their degeneracy (= 2 for 2$p$ photoemission) and spin-orbit splitting (= 35.6 eV) is also kept similar in every case to minimize adjustable parameters for better reliability.

In the bulk As 2$p$ spectrum, there are two distinct features in each of the spin-orbit split feature as marked by `1' and `2' in Fig. \ref{As2pFit}(b) for As 2$p_{3/2}$ spectrum. Therefore, we used two Voigt function for each spin-orbit split features as shown in Figs. \ref{As2pFit}(a) and (b). The binding energy of `1' and `2' are 1321.9 eV and 1322.7 eV, respectively. As3 atoms forming the cis-trans chain are monovalent and weakly coupled to the adjacent Ce layers while  As1/As2 are trivalent\cite{demchyna}. Thus, the lower binding energy peak, `1' in the figure can be attributed to As1/As2 2$p$ photoemission and the feature `2' appears due to As3 2$p$ photoemission signal. In addition, we observe an intense peak at about 1338 eV and is attributed to the plasmon excitation feature; signature of such loss features are also observed in the other core level spectra.

In the surface spectra, we observe additional features; distinct features are marked in Fig. \ref{As2pFit}(d) by `3' and `4'. We simulated the surface spectrum following the same method as done for bulk spectrum and the binding energy of the features `1' and `2' are found to be similar to the bulk case. The binding energies of the features `3' and `4' are 1323.7 eV and 1325.6 eV, respectively. The features `1' and `2' in Fig. \ref{As2pFit}(d) are very similar to those in Fig. \ref{As2pFit}(b) and can be attributed to As1/As2 and As3 layers, respectively. The enhancement of As3 intensity in the surface spectra is understandable as the terminated surface is found to be As3-layer. The features `3' and `4' appear at binding energies higher than the binding energy found even in metallic As\cite{wagner} indicating positive As valency which is unexpected in CeAgAs$_2$. The binding energy of `4' is exactly equal to divalent As in AsO\cite{bhal}. The feature, `3' can be due to monovalent As as often found at the surface of various other materials \cite{ZrB12-1,ZrB12-2}. Thus, we attribute the features `3' and `4' to the oxides of As formed due to the surface oxygen discussed later in the text.The total intensity of the features `3' and `4' is found to be $<$10\% of intensity of the other peaks.
The features at higher binding energies are due to the plasmon excitations.
	
\begin{figure}
\centering
\includegraphics[width=0.5\textwidth]{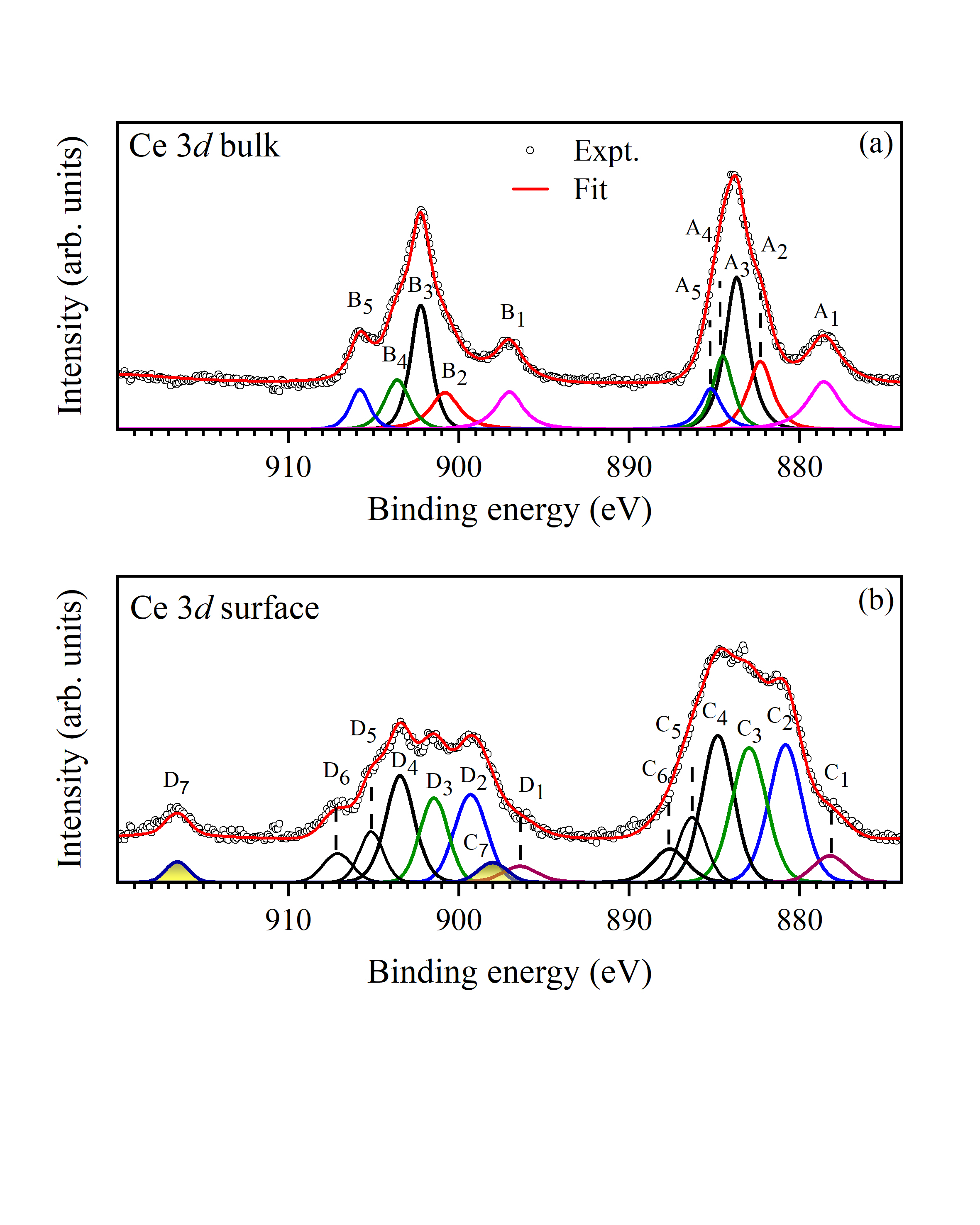}
\vspace{-16ex}
\caption{Ce 3$d$ (a) bulk and (b) surface spectra. Lines represent the simulated spectra using voigt function for each of the distinct features. Lower panel show the component features.}
\label{Ce3dFit}
\end{figure}

The surface and bulk contributions in the Ce 3$d$ core level spectra are extracted using 6 keV and 2.5 keV data following the same method discussed above. The extracted bulk and surface Ce 3$d$ features are shown by the black open circle in Fig. \ref{Ce3dFit}(a) and (b), respectively. Multiple peak structured spectra are observed for both surface and bulk Ce 3$d$ contributions. In the bulk spectra shown in Fig. \ref{Ce3dFit}(a), we observe 5 distinct signature of peaks in each of the spin-orbit split signal. In the 3$d_{3/2}$ case, they are denoted by B$_1$, B$_2$, B$_3$, B$_4$ and B$_5$. The corresponding peaks in the 3$d_{5/2}$ signal are represented by A's. In photoemission, the spectral function represents the photoemission response manifested in the excited state in the presence of photo-hole. The presence of core hole makes the final state Hamiltonian different from the initial state Hamiltonian if there is finite electron-electron correlation. Thus, the photoemission signal will contain features corresponding to the transition to each of these final states which are different from the ground state; this is known as the final-state effect. In the Gunnarsson and Schonhammer's (GS) description \cite{gun1, gun2}, based on single impurity Anderson model for Ce 3$d$, the final states can be expressed in terms of electronic configurations such as (i) $well-screened$ final state, $f_2$ - the core-hole is screened via hopping of a valence electron to the Ce 4$f$ level of the photoemission site. The presence of this feature depends on the hybridization of the 4$f$ states with the valence band \cite{nickelates_kbm, RMP_Fujimori}. (ii) $Poorly-screened$ final state, $f_1$ corresponds to the case where the core holes are not screened. (iii) $f_0$ final state where the 4$f$ electron at the photoemission site has lost its local character due to the hopping to the ligand bands and/or conduction band in the presence of finite hybridization between them. Such descriptions are found to work well in 4$f$-based compounds.

The 4$f$ electrons entangled with the conduction electrons due to Kondo coupling gain itineracy. The corresponding feature in the Ce 3$d$ spectrum will appear close to the $f_0$ state as the Ce 4$f$ electron is no longer a local electron. The intensity of the Kondo singlet states enhances with the decrease in temperature. The binding energies of $f_2$, $f_1$ and $f_0$ states appear in increasing order which corresponds to the effective local charge at the photoemission site.

Features observed in the Ce 3$d$ spectra can be attributed to various final states discussed above. We have simulated the experimental spectra using a set of Voigt functions keeping the branching ratio of the intensity of the spin-orbit split features fixed to the ratio of degeneracy (= 1.5) and the spin-orbit splitting of 18.6 eV. The best fits are shown by red lines superimposed over the raw data. The constituent peaks are shown in the lower panel. The bulk Ce 3$d$ spectrum shown in Fig. \ref{Depth}(a) resembles well the bulk spectra of the CeCuSb$_2$\cite{sawani}. Considering the binding energies of the peaks\cite{sawani}, the features A$_2$-A$_5$ in 3$d_{5/2}$ signal (B$_2$-B$_5$ in 3$d_{3/2}$ signal) are attributed to the multiplets of $f_1$ and $f_2$ final states. The low energy feature, A$_2$ (B$_2$ in 3$d_{3/2}$ feature) corresponds to the well screened, $f_2$ state. In addition, there are low energy intense features A$_1$ and B$_1$. Signature of such features are found in some systems\cite{sawani,Ram, ZR}, where the hole created in the conduction/legand bands due to the hopping of an electron to the core hole site, forms a singlet with other holes and/or has highly itinerant character that allows the valence hole to propagate far away from the photoemission site to lower the energy further. The intensity of these features appear to be quite large in this material.

The surface spectrum is shown in Fig. \ref{Ce3dFit}(b); features are represented by C's and D's for 3$d_{5/2}$ and 3$d_{3/2}$ spectral regions, respectively. The relative intensities of the features are different from those in the bulk spectrum. In addition, there are distinct additional peaks at higher binding energies indicating presence of tetravalent Ce at the surface\cite{jan, Zhang}. We observe intense $f_0$ peak (D$_7$ for Ce 3$d_{3/2}$ signal) in the surface spectrum which is almost absent in the bulk. On the other hand, the intensity of C$_1$ and D$_1$ is weaker than those in the bulk spectrum. It appears that Ce-layers closer to the surface can host higher valency compared to those in the bulk. Moreover, the itineracy at the surface may be weaker arising from surface termination induced band narrowing that led to reduced intensity of the lowest energy feature.


\begin{figure}
\centering
\includegraphics[width=0.5\textwidth]{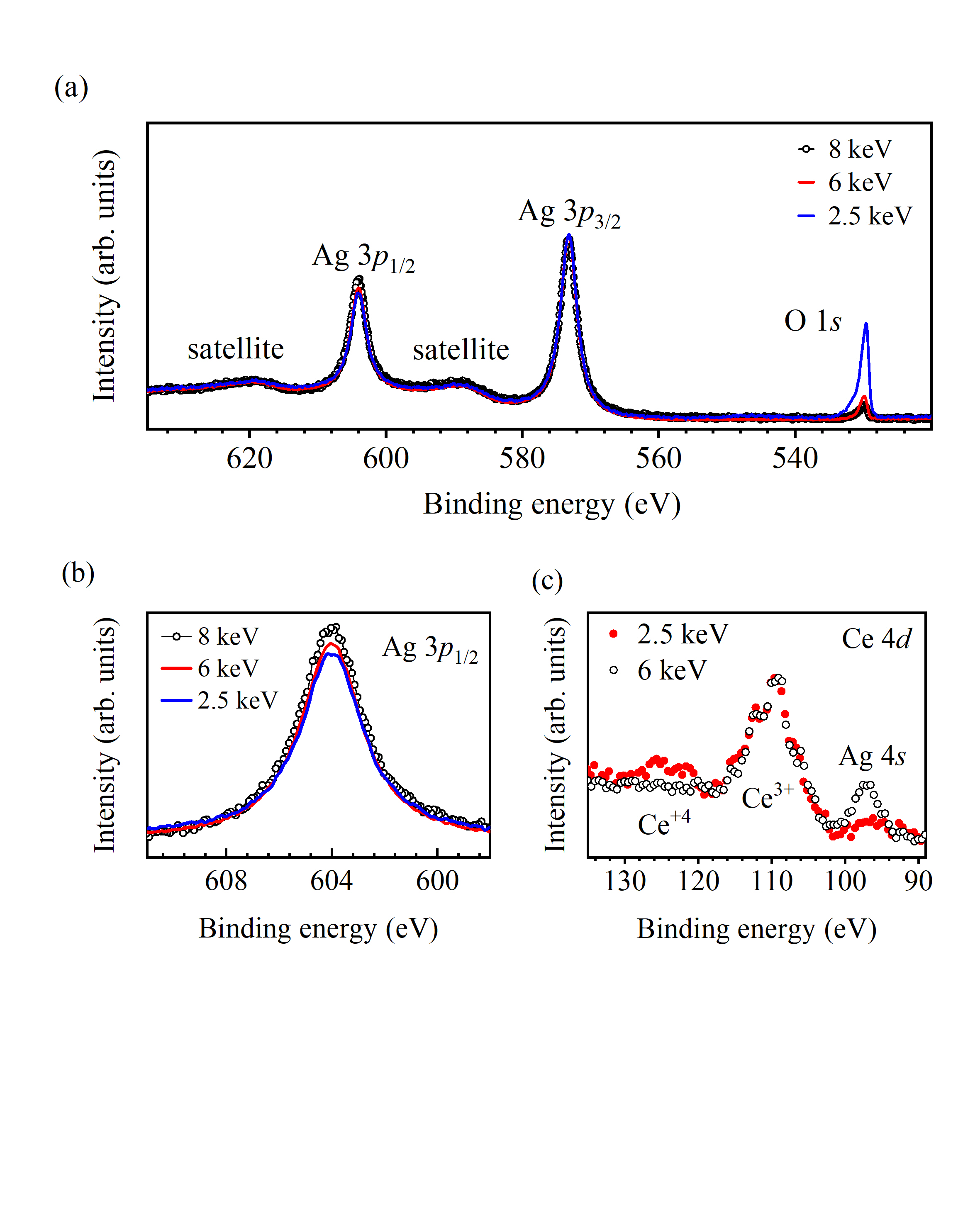}
\vspace{-18ex}
\caption{(a) Ag 3$p$ collected using 8 keV (open circles), 6 keV (red line) and 2.5 keV (blue line) photon energies. Ag 2$p$ spectra show signature of satellite features. 2.5 keV data show significant O 1$s$ signal which become weak at higher photon energies. (b) Ag 3$p_{1/2}$ feature at different photon energies shown in an expanded energy scale exhibiting enhancement of intensity with photon energy relative to 3$p_{3/2}$ spectral intensity. (c) Ce 4$d$ spectra at 2.5 keV (closed circles) and 6 keV (open circles). Experimental data show signature of Ce$^{4+}$ in the surface spectra and large Ag 4$s$ signal in the bulk spectra.}
\label{Ag2pO1s}
\end{figure}

In order to investigate the above observations further, we have studied the Ag 2$p$ and O 1$s$ spectra collected using 2.5 keV, 6 keV and 8 keV photon energies. The experimental data are shown in Fig. \ref{Ag2pO1s}(a). While the lineshape of Ag 2$p$ spectra exhibit quite similar behavior at all the photon energies, the O 1$s$ intensity is significantly strong in the 2.5 keV data (blue line in the figure). Oxygen 1$s$ intensity becomes negligibly small at 6 keV and reduces further in the 8 keV data. This suggests that the surface of the sample, though cleaved in ultrahigh vacuum, contains small amount of oxygen. We have estimated the intensity of the surface oxygen assuming atomic cross-section of various core levels and their depth from the surface layer. We find that oxygen coverage at the surface is less than 10\%. We notice that the total intensity of the features `3' and `4' in the surface As 2$p$ spectrum shown in Fig. \ref{As2pFit}(d) is also close to 10\% of the total As intensity. This explains the observation of additional higher binding energy features in the As 2$p$ and Ce 3$d$ core level spectra.

Ag 2$p$ spectra show two spin-orbit split features at 572 eV and 604 eV along with two intense satellite features which are about 16 eV away from the main peak and can be attributed to the plasmon excitations as also observed in other core level spectra. Interestingly, a normalization of the spectra by the intensity of the 2$p_{3/2}$ signal show enhancement of intensity of the 2$p_{1/2}$ signal with the increase in photon energy as shown in Fig. \ref{Ag2pO1s}(b). Such enhancement suggests increase of the branching ratio beyond the degeneracy of the atomic description. Such enhancement may occur due to enhancement of the orbital moment in the solid arising from uncompensated electric field in the bulk as also observed in other systems \cite{BiPd}.

Ce 4$d$ spectra shown in Fig. \ref{Ag2pO1s}(c) exhibit a peak at about 110 eV binding energy in the 6 keV data suggesting presence of Ce$^{3+}$ in the bulk. The 2.5 keV data show an additional intensity at around 122 eV which can be attributed to the Ce$^{4+}$ signal\cite{Dudric}. In addition, we observe intense Ag 4$s$ peak at about 96 eV in the 6 keV data which becomes very weak in the 2.5 keV data indicating that the depth of the Ag-layers from the surface is larger than that for the Ce-layer. This observation along with the observation in Fig. \ref{structure} experimentally establish that As3 layer is the terminated surface layer of our cleaved sample.

\begin{figure}
\centering
\includegraphics[width=0.5\textwidth]{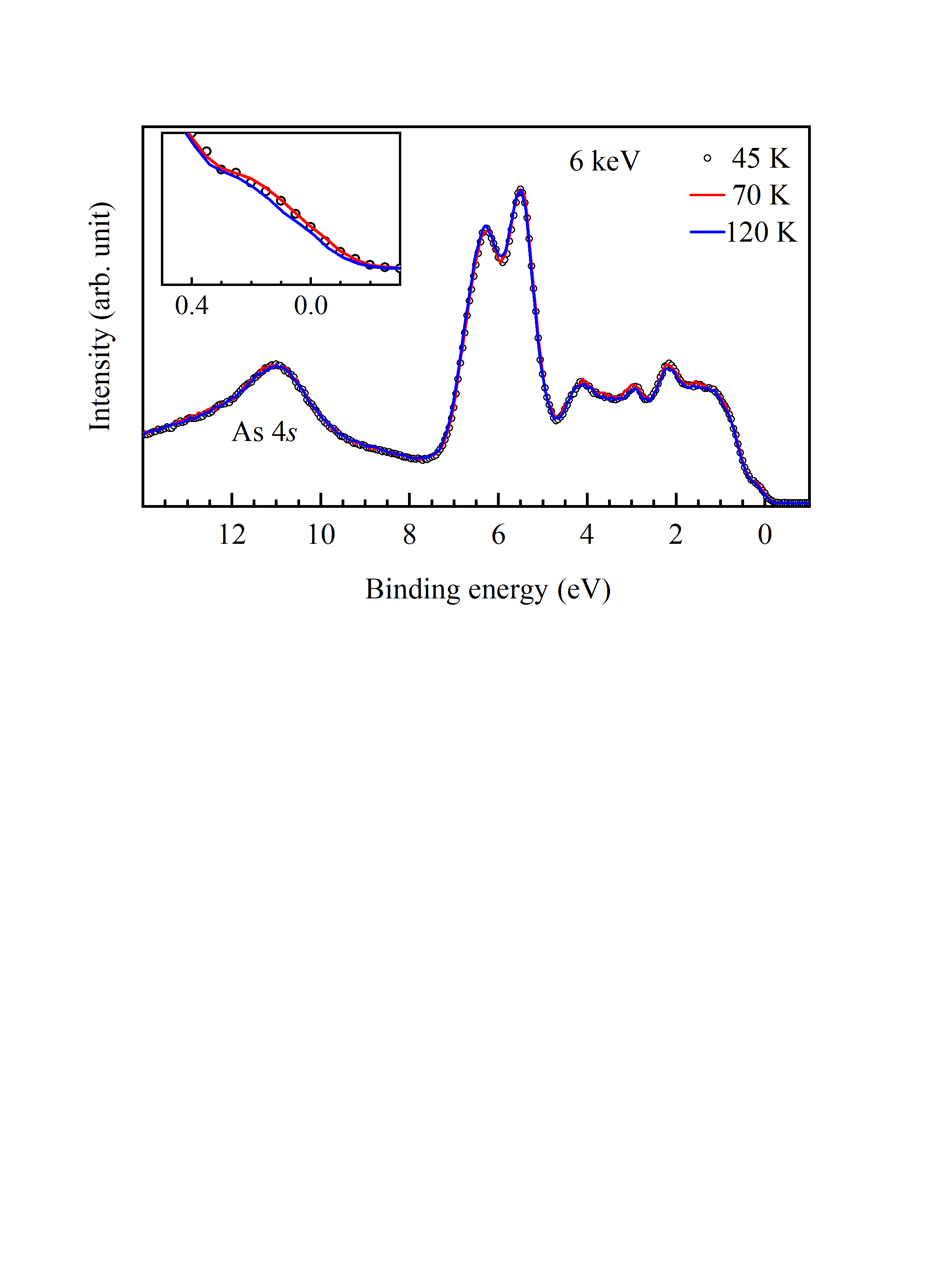}
\vspace{-48ex}
\caption{Valence band spectra collected at 45 K (symbols), 70 K (red line) and 120 K (blue line) using 6 keV photon energy. Inset shows the near Fermi level region in an expanded scale exhibiting depletion of intensity with the increase in temperature.}
\label{FigTempVB}
\end{figure}

We now turn to the temperature dependence of the electronic structure; the experimental data for the valence band spectra collected at different temperatures are shown in Fig. \ref{FigTempVB}. Here we have shown the 6 keV data which represent essentially the bulk electronic structure and also has the highest energy resolution in our measurements. Interestingly, all the spectra collected at 45 K, 70 K and 120 K superimpose over each other excellently well. The spectral region close to the Fermi level is shown in an expanded scale in the inset of the figure. There is a depletion of intensity at the Fermi level with the increase in temperature. Observation of such a change, though small, may be important as the photoemission cross section for the Ce 4$f$ states relative to other valence states is very small at this photon energy. Thus, the observation of this depletion of intensity with the increase in temperature may be a signature of the Kondo effect observed in the transport and magnetic behavior.

\begin{figure}
\centering
\includegraphics[width=0.5\textwidth]{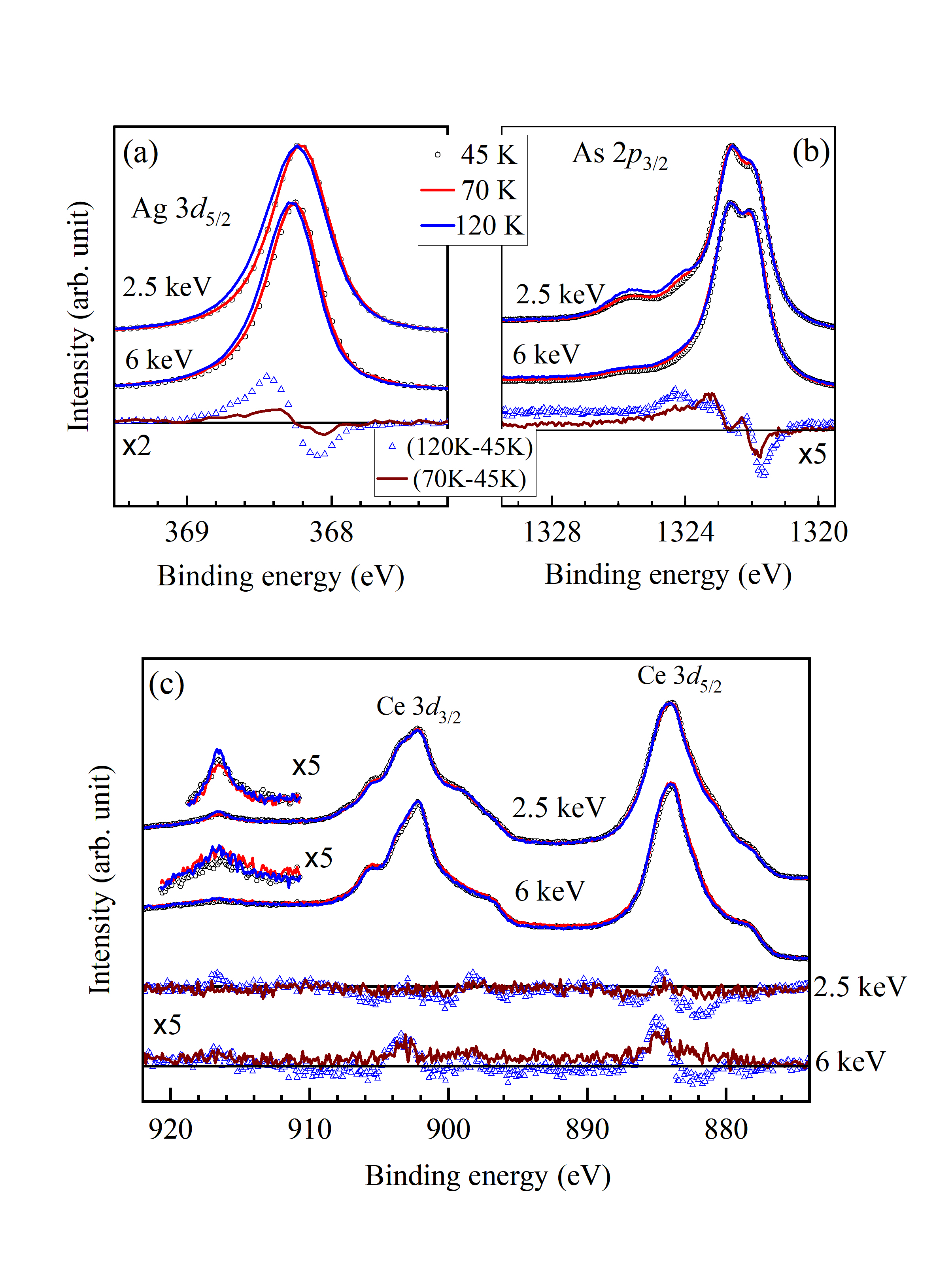}
\vspace{-6ex}
\caption{(a) Ag 3$d_{5/2}$, (b) As 2$p_{3/2}$ and (c) Ce 3$d$ spectra collected at 45 K (open circles), 70 K (red line) and 120 K (blue line) using 2.5 keV (upper panel) and 6 keV (lower panel) photon energies. (c) Ce 3$d$ spectra collected at these temperature and 2.5 keV photon energy. The lower panels in each of the figures show the difference spectra for (120K-45K) (open up-triangles) and (70K-45K) (solid line) which show the spectral changes with heating.}
\label{FigTempCore}
\end{figure}

Ag 3$d_{5/2}$, As 2$p_{3/2}$ and Ce 3$d$ spectra collected at 45 K, 70 K and 120 K using 2.5 keV and 6 keV photon energies are shown in Fig. \ref{FigTempCore}. All the spectra are normalized by the intensity of the most intense feature. Ag 3$d_{5/2}$ spectra shown in Fig. \ref{FigTempCore}(a) exhibit a gradual shift towards higher binding energies with the increase in temperatures in both the cases. This is evident in the difference spectra shown in the bottom panel of the figure for 6 keV data where the difference of the higher temperature spectrum with respect to the 45 K data are shown. The increase in temperature exhibits a spectral weight transfer towards higher binding energies in the difference spectra. Such a shift of the peak suggests that the eigenenergy of the well screened final state increases with the increase in temperature. This may happen due to the heating induced increase in bondlength which reduces covalency thereby enhancing the effective Madelung potential at the Ag-sites. The hopping interaction strength may also reduce leading to a less efficient screening of the core holes.

As 2$p_{3/2}$ features exhibit interesting scenario. A normalization of the spectra by the intensity of the As3 2$p_{3/2}$ peak at 1322.7 eV in the 2.5 keV data show gradual enhancement of intensities of the peaks at 1323.7 eV and 1325.6 eV which might occur due to higher degree of oxidation of the surface As atoms. The spectral changes in the 6 keV data suggest that with the increase in temperature the intensity around 1322 eV reduces gradually with consequent increase beyond 1323 eV; this is shown with clarity in the difference spectra in the lower panel of the figure. It is clear that the intensity of the lowest energy final state reduces with the increase in temperature giving rise to an emergence of intensities at higher binding energies which corresponds to higher energy eigenstates of the final state Hamiltonian.

The temperature evolution of the Ce 3$d$ states are analysed in Fig. \ref{FigTempCore}(c). In the wide energy scale, the spectra collected at different temperatures look quite similar with subtle changes close to the most intense features for both the photon energies, 6 keV and 2.5 keV. A closer observation shows subtle differences in the intensity of the spectral features. The $f_0$ peak at 916 eV is shown in an expanded scale exhibiting similar intensities (integrated area corresponding to the feature) at all the temperatures. We subtracted the 45 K data from the higher temperature data; the difference spectra are shown in the lower panel of the figure. The spectral changes are quite small in the 2.5 keV data. In the 6 keV case which represent primarily the bulk electronic structure, the intensity around 882 eV corresponding to $f_2$ feature appear to reduce along with an enhancement around 885 eV where $f_1$ intensities appear; a similar scenario is observed in the 3$d_{3/2}$ spectral region too. These results suggest that the core hole screening is more efficient at lower temperatures presumably due to enhanced coupling of the 4$f$ states with the conduction electronic states at low temperatures.

\section{Conclusion}
	
In conclusion, we have studied the electronic structure of an antiferromagnetic Kondo lattice system CeAgAs$_2$ employing high-resolution hard $x$-ray photoemission spectroscopy. The DFT results provide a reasonable description of the features in the valence band spectra. The experimental spectra collected at different photon energies suggests that the cleaving occurs at the As3 layer. The depth-resolved data show significant surface-bulk differences in the Ce and As core level spectra. The As 2$p$-bulk spectra show two distinct features corresponding to As3 and As1/As2 atoms. As3 peaks appear at higher binding energies compared to As1/As2 peaks suggesting lower negative valency of As3 formed in cis-trans structure compared to As layers strongly coupled to Ce and Ag layers.

The Ce 3$d$ spectra exhibit multiple features due to the final state effects arising from strong Ce-As hybridization and electron correlation. The bulk Ce 3$d$ spectra exhibit multiplet structure similar to other CeTX$_2$ (T=Cu, Ag; X= As, Sb) compounds suggesting close to trivalency of bulk Ce. In addition to the well-screened and poorly-screened features, we observe a feature at lower binding energy which becomes intense in the bulk spectra. The presence of such low energy final states provides an evidence of high degree of itineracy of the valence hole and/or singlets as found in other systems \cite{Ram, ZR}. The $f_0$ peak, a signature of Kondo effect is almost absent in the bulk Ce 3$d$ spectra though the materials show Kondo-like properties; the $f_0$-intensity is finite in the surface spectra. However, the valence band spectra show cooling induced enhancement of the spectral intensity close to the Fermi level as expected in a Kondo system. The core level spectra show significant spectral redistribution with heating; there is a spectral weight transfer from well screened feature to higher binding energies. This suggests that the screening of the core hole is more efficient in the bulk electronic structure and at low temperatures. It appears that the spectral renormalization arising from temperature induced complex evolution of inter- and intra-layer covalency in the presence of strong electron correlation plays the key role in deriving the exotic properties of the system.

\section{Acknowledgements}
Authors acknowledge the financial support under India-DESY program and Department of Atomic Energy (DAE), Govt. of India (Project Identification no. RTI4003, DAE OM no. 1303/2/2019/R\&D-II/DAE/2079 dated 11.02.2020). K.M. acknowledges financial support from BRNS, DAE, Govt. of India under the DAE-SRC-OI Award (grant no. 21/08/2015-BRNS/10977).

\end{document}